# AN EPICS TO TINE TRANSLATOR

Zoltan Kakucs, Phil Duval, Matthias Clausen
DESY Hamburg, Germany

Abstract

Accelerator control at DESY has in the past been hampered by the 'many-control-systems' syndrome, where different subsystems were controlled by completely different means offering no possibility of intercommunication. This was particularly true in HERA during the commissioning phase. Today, practically all subsystems of HERA are controlled by TINE [1]. Important exceptions include the Proton Vacuum (DOOCS [2]), cryogenics control (D3), and the super-conducting electron RF cavities and the power and cooling subsystems, the latter two of which are controlled by EPICS IOCs. A step toward integrating the EPICS IOCs into the HERA mainstream has been taken in that an EPICS to TINE translator process has been written, which runs directly on the EPICS IOC and offers a TINE view of the hardware control to the rest of the control system. An EPICS IOC can then be controlled via channel access as before, and in addition via TINE protocol. The server module of the translator resides in each one of the system controllers along with that controller's portion of the distributed EPICS staff and database. This in effect renders the EPICS IOC into a bi-lingual server, and highlights the principal difference between a 'translator' and a 'gateway'. The EPICS PV names registered as TINE devices, the fields of the records are registered as TINE properties. The server is able to send (receive) any value to (from) any client TINE application, in any of the TINE modes of data acquisition. The multiple instances of the server in a control system respond to a request for data by searching for the registered device. If a server has the requested data in the local database, it responds to the request and sends the data to the client. This allows any of an arbitrary number of data-using clients to receive data from any of an arbitrary number of data-supplying servers without the client needing to know the location or other attributes of the data. The result is an easily scaleable translator system. Details of this translator are presented in this paper.

## 1 GENERAL INFORMATION

### 1.1 A few words about EPICS

EPICS (Experimental Physics and Industrial Control System) is a software toolkit with which Application Developers can create a control system. EPICS can work under different operating systems such as the real-time OS VxWorks, Solaris as well as Linux. Physically, it is a flat, client/server based architecture of front-end controllers and operator workstations that communicate via TCP/IP and UDP. Some basic components are:

- OPI (Operator Interface): can be a UNIX or NT-based workstation or PC running various EPICS tools.
- IOC (Input Output Controller): can be a VME/VXI based chassis containing a Motorola 68K or a PPC processor with different VME I/O modules for various signals and field bus accesses
- LAN (Local Area Network): this connects the IOCs and OPIs using the network transparent TCP/IP based communication software component of EPICS namely Channel Access (CA)

### 1.2 Channel Access

Channel Access (CA) is the EPICS network protocol based on the Client/Server model. Each IOC runs its own local CA-Server task, which establishes connections. At the other end, the CA-client provides access to the process databases in other IOCs and uses TCP for data transport and UDP for connection management. External units access the database via CA. Additionally the internal units can access a record in the database directly through the database access routines.

### 1.3 A few words about TINE

TINE (Threefold Integrated Networking Environment) runs on any number of platforms including such legacy platforms as MSDOS, VAX VMS as well as all brands of UNIX, all brands of WINDOWS and most importantly for our purposes here VxWorks. Like EPICS, TINE is a distributed system, but with central services such as naming services, alarm services, archive services, etc.

TINE servers also have a built-in local alarm server, local history server, and a number of stock query services. Data can be transferred over a number of different architectures and protocols, and there is no practical limit to the size of a data block transmitted from a server. The goal is to leverage the functionality behind TINE and the process control and database structure behind EPICS, and at the same time not to disturb existing CA communication channels.

## 2 HISTORY AND FIRST STEPS

### 2.1 EPICS data to TINE in the past

There were two different ways to get EPICS data into other systems in the past, but they're based on the same connection method, namely CA. One method involved using the EPICS visualization tools. Another was to use the built-in CA subroutines. The last one linked the CA libraries to any third party program, for instance a TINE based application. Some requirements would still be unrealized at the Client-Side as they are specific to EPICS.

Disadvantages:
- Unfriendly update of distributed CA libraries and DLL
- Special VBA using CA functions
- No naming service available
- Low priority clients consume resources in critical machines

### 2.2 First Ideas

The main goal was to integrate the EPICS IOC-s into the HERA mainstream. Considering the advantages of EPICS, namely modularity and customization, the preferable way should be an additional server running directly on the EPICS IOC. The server module should reside in each one of the system controllers along with that controller's portion of the distributed EPICS context. This would provide the TINE view of the hardware control to the rest of the Control System.

### 2.3 Requirements

- provide excellent performance without disturbing the real-time control loops in and between subsystems.
- offer a maximum level of functional flexibility without additional programming
- each server module has to have its own "mapped" record list
- under all circumstances the control via channel access must remain as before
- the server module of the translator must reside in each one of the system controllers along with that controller's portion of the distributed EPICS staff and database
- the translator has to fit seamlessly into the TINE systematic (Alarm, Archive, Naming, Permit,…)
- each server module should map the record names to TINE registered names
- building set of process variable names mapped to common device or property names, called "composites"

## 3 FUNCTIONALITY

### 3.1 Naming convention

The EPICS ASCII database is the heart of an IOC, presenting the records. All record names are unique across all IOC-s attached to the same TCP/IP subnet. The form is <record name>[.<field name>]. The access to the database is via the channel or database access routines. The record name is up to 28 chars long, the field can be up to 4 chars long.

The TINE naming convention follows the hierarchy <device context>/<device server>/<device name> (each 16 characters long) and allows data flow via <device property> (32 characters long). One obstacle was to systematically map the EPICS record names into the TINE name space. Logically, one tries to make the correspondence between the EPICS field name (typically "DESC","VAL", etc.) and a TINE property (this fits with 28 characters to spare!) and the EPICS record name with the TINE device name. However we have the immediate problem of mapping 28 characters onto 16. Of course, pieces of EPICS record names frequently fit into the TINE context and TINE device server names. However there is no systematic behind this. Our solution was to make use of feature of TINE, where so-called "long device names" (32 characters) can be used in conjunction with "short" property names (16 characters), which is in fact the case we have.

### 3.2 Data requiring

If a TINE client requires data, the translator follows the way described below:
- reconstruct the process variable name, needed for the database access
- search for the record name using directly the database access layer.
- locally convert the required data corresponding to the TINE client requested data type format
- send back the data to the client
- if the requested data is a composite, then loop over the records making up the composite.

The EPICS IOC doesn't need to know the location or other attributes of the data. The request is coming via

TINE. Name servicing is built into the TINE environment, so that this translator can fully use it.

### 3.3 Security

For more security it is possible to enable the CA layer instead of the database access layer using a software switch at startup time. Indeed would mean for all access a double security layer would be employed, i.e. both TINE security and CA security.

### 3.4 Data conversion

The data type conversions are performed in the server, using EPICS standard data types. All converting functionality was possible without major changes in EPICS or TINE code.

### 3.5 Composites

Composites are set of arranged TINE devices identified through collective names. The configuration files setting up this composites contains the collective name and the members of each composite device. Thus an object-view of the IOC with human readable property names and descriptions can be offered to the clients (e.g. "ORBIT" instead of a series of he:adc:wl244:orbit_ai)

### 3.6 Server management

The server has a built-in management system. It can be started at any time after EPICS is running on any system. All components, such as the library and the objects can be downloaded independently from EPICS code. During the initial tests it was important to be able to stop the translator without interfering the running EPICS core. The actual version has a built-in starting, stopping and restarting mechanism without affecting the CA communication.

### 3.7 Tests

Tests have proved the stable monitoring up to 2500 channels by one IOC

## 5 SUMMARY AND CONCLUSIONS

- Possibility to communicate between two different subsystems
- Runtime integrability of any EPICS running subsystem
- Transparency of the EPICS database given in the TINE systems
- Integrated TINE server source code into the legacy EPICS tree
- Easy EPICS version control, without distributing the CA libraries
- No need of special TINE Application to access EPICS data
- Avoiding subnet dependency of CA Protocol by using the TINE
- Easy-to-use built-in naming service in TINE

Learning to work together effectively was a significant ingredient in the success of this work. It became evident that research programs needed to be able to do independent system development to meet the requirements of their projects. The clean interfaces that make many of the extensions simple to add are a portion of this success.